\documentclass[a4paper,12pt]{article}
\usepackage{amsmath}
\usepackage{amsfonts}
\usepackage{amssymb}
\usepackage{graphicx}

\title{{\textsc{\begin{LARGE}Gravitons in Kaluza-Klein Theory\end{LARGE}}}}
\author{\textbf{$\hbox{V H Satheesh Kumar}^{\star}$} and \textbf{{$\hbox{P K Suresh}^{\dagger}$}} \\ \textit{School of Physics, University of Hyderabad} \\ \textit{Hyderabad 500 046, India.} \\ $^\star$ {\begin{small}\texttt{vhsatheeshkumar@yahoo.co.uk}\end{small}}, $^\dagger$ \begin{small}\texttt{pkssp@uohyd.ernet.in}\end{small}}
\date{}

\begin{document}

\maketitle

\begin{abstract}
This is a pedagogical introduction to original Kaluza-Klein theory and its salient features. Most of the technical calculations are given in detail and the nature of gravitons is discussed. 
\end{abstract}
\newpage
\section{Introduction}
The current trend in modern theoretical physics is to search for a theory which provides a unification of gravity with the other fundamental interactions of nature. One of the early possibilities for such a unification of the then known interactions i.e., gravity and electromagnetism, was suggested by Kaluza[1] and Klein[2]. Gravity and electricity are indeed very similar in many ways, but the relation between them is not nearly as straight forward as is suggested by the fact that in classical physics they both are governed by inverse square laws. Relativistically, for instance, the field equations of electromagnetism i.e., Maxwell's equations are linear; while the field equations of gravitational field i.e., Einstein's equations are highly nonlinear. Obviously it is quite a difficult task to bring them under one roof without some exotic idea. Kaluza and Klein showed within the context of five dimensional extension of Einstein's theory of general relativity how both electromagnetism and gravity could be treated on a similar footing -- in the sense that both were described as parts of the five-dimensional metric.

But historically, it was Gunnar Nordstr\"om[3] who brought the idea of extra spacial dimension into physics. In 1914, he discovered that one could unite the physics of electromagnetism and gravity by postulating the existence of a fourth spatial dimension. He considered the five dimensional vector potential which can be written in its (4+1) form as,
\begin{equation}
A_{\hat\mu} = A_\mu + \phi
\end{equation} 
with $A_\mu$ as four vector potential of electromagnetism and scalar $\phi$ identified as gravitational field. Here onwards all hatted quantities $(\hat{\mu}=0,1,2,3,4)$ are five-dimensional and all unhatted quantities $(\mu=0,1,2,3)$ are four-dimensional. 

However, Nordstr\"om worked with a scalar theory of gravity, not with a tensor theory like general relativity, which had not been published yet. When Einstein published his theory of general relativity, Nordstrom abandoned his own approach. But the idea of extra dimension was here to stay! A recent non-technichal review of extra dimensional theories can be found in [4].

In the coming sections we give the complete pedagogical account of the original Kaluza-Klein theory and discuss its  salient features. Here, we are not investigating many other theories built upon Kaluza-Klein's idea, for which a reader can refer to very good reviews [5].
\section{Kaluza's Fifth Dimension}
An early proposal to unite general relativity  and classical electrodynamics was given by Theodor Franz Eduard Kaluza[1] in 1921. He showed that the gravitational and electromagnetic fields stem from a single universal tensor and such an intimate combination of the two interactions is possible in principle, with the introduction of an additional spacial dimension. Although our rich physical experience obtained so far provides little suggestion of such a new spacial dimension, we are certainly free to consider our world to be four dimensional spacetime of the bigger five dimensional spacetime. In this scenario, one has to take into account the fact that we are only aware of the spacetime variation of state-quantities, by making their derivatives with respect to the new parameter vanish or by considering them to be small as they are of higher order. This assumption is known as the \textit{cylindrical condition}. 

The five dimensional line element is given by 
\begin{equation}
d\hat{s}^2 = \hat{g}_{\hat{\mu}\hat{\nu}}(x^\mu,y)d\hat{x}^{\hat{\mu}} d\hat{x}^{\hat{\nu}}
\end{equation} 
with $y$ as the additional spatial coordinate. The five dimensional metric can be expressed as,
\begin{equation}
\hat{g}_{\hat{\mu}\hat{\nu}}=
\left( \begin{array}{cc}
g_{\mu\nu} & g_{\mu 5}\\
g_{5 \nu} & g_{5 5}\\
\end{array}\right)
\end{equation} 
 
Once we have a spacetime metric, like in standard general relativity we can construct the Christoffel symbols $\Gamma^\mu_{\nu \rho}$, the Riemann-Christoffel curvature tensor $R^\mu_{\nu \rho \sigma}$, the Ricci tensor $R_{\mu \nu}$, the curvature invariant $R$ and then the field equations. Now we compute these quantities one by one for the five dimensional metric.
\subsection{Christoffel Symbols}
By virtue of the cylindrical condition, the Christoffel symbols of the first kind for $\hat{g}_{\hat{\mu}\hat{\nu}}$ turn out be,
\begin{eqnarray}
2\Gamma_{\lambda\mu\nu}&=&g_{\lambda \mu , \nu}-g_{\mu \nu, \lambda}-g_{\nu \lambda , \mu} \; \mathtt{(40\, equations\, as\, in\, 4D\, GR)}\\
2\Gamma_{\lambda\mu 5}&=&-g_{5 \lambda, \mu}-g_{5 \mu, \lambda}\;\;\;\;\;\;\;\;\;\; \mathtt{(16\, equations)}\\
2\Gamma_{5\mu\nu}&=&g_{5 \mu , \nu}-g_{5 \nu, \mu} \;\;\;\;\;\;\;\;\;\;\;\;\;\mathtt{(10\, equations)}\\
2\Gamma_{5 5\nu}&=&g_{5 5, \nu}\;\;\;\;\;\;\;\;\;\;\;\;\;\;\;\;\;\;\;\;\;\;\;\;\; \mathtt{(4\, equations)}\\
2\Gamma_{5\mu 5}&=&-g_{5 5 , \mu} \;\;\;\;\;\;\;\;\;\;\;\;\;\;\;\;\;\;\;\;\;\; \mathtt{(4\, quations)}\\
2\Gamma_{5 5 5}&=&0 \;\;\;\;\;\;\;\;\;\;\;\;\;\;\;\;\;\;\;\;\;\;\;\;\;\;\;\;\; \mathtt{(1\, equation)}.
\end{eqnarray} 
As we expect for a five dimensional metric, it has 75 independent Christoffel symbols. Now we set 
\begin{eqnarray}
g_{5\mu} &=& 2 \alpha A_\mu\\
g_{55} &=& 2 \phi
\end{eqnarray} 
so that $\hat{g}_{\hat{\mu}\hat{\nu}}$ becomes the gravitational tensor potential framed by the electromagnetic four-potential $A_\mu$ and scalar field $\phi$. Now, the above equations take the form,
\begin{eqnarray}
\Gamma_{5\mu\nu}&=&\alpha (A_{\mu, \nu}- A_{\nu, \mu})=\alpha F_{\mu \nu}\\
\Gamma_{\mu\nu 5}&=&-\alpha (A_{\mu, \nu}- A_{\nu, \mu})=-\alpha \Sigma_{\mu \nu}\\
\Gamma_{5 5 \mu}&=&-\Gamma_{5 \mu 5}=\phi,\mu.
\end{eqnarray} 

Now going back a step, the first 40 Christoffel symbols give us the standard general relativity and the remaining 35 Christoffel symbols can be related by $F_{\mu \nu}$ and $\phi$ through the relations
\begin{equation}
(\Gamma_{lmn}+\Gamma_{mnl}+\Gamma_{nlm})_{,k}=\Gamma_{klm,n}+\Gamma_{kmn,l}+\Gamma_{klm,n}.
\end{equation} 
 Thus, we get 
\begin{eqnarray}
F_{\mu \nu, \lambda}+F_{\nu \lambda , \mu}+F_{\lambda \mu , \nu}&=&0\\
\phi_{,\mu \nu}&=&\phi_{,\nu \mu}.
\end{eqnarray} 
\subsection{Riemann-Christoffel Tensor}
As a next step in our investigation, we compute the Riemann-Christoffel tensors for $\hat{g}_{\hat{\mu}\hat{\nu}}$, under the approximation that $\hat{g}_{\hat{\mu} \hat{\nu}}$ differs slightly from Minkowski metric $\hat{\eta}_{\hat{\mu} \hat{\nu}}$, that is,
\begin{equation}
\hat{g}_{\hat{\mu} \hat{\nu}} = \hat{\eta}_{\hat{\mu} \hat{\nu}}+ \hat{h}_{\hat{\mu} \hat{\nu}} 
\end{equation}  where $|\hat{h}_{\hat{\mu} \hat{\nu}}|\ll 1.$
So, under the weak field approximation, we can write $\Gamma^{\hat{\lambda}}_{\hat{\mu} \hat{\nu}}=-\Gamma_{\hat{\lambda} \hat{\mu} \hat{\nu}}$. In five dimensions, there will be $5^5=3125$ components for $R^{\hat{\mu}}_{\hat{\nu} \hat{\lambda} \hat{\sigma}}$, but because of the many symmetries they obey, the number will be redused to only 50 independent components.
\begin{eqnarray}
R^\mu_{ \nu \lambda \sigma}&=&\Gamma^\mu_{ \nu \sigma, \lambda}-\Gamma^\mu_{ \nu \lambda, \sigma}+\Gamma^{\mu}_{\rho \lambda}\Gamma^\rho_{\nu \sigma}-\Gamma^{\mu}_{\rho \sigma}\Gamma^\rho_{ \nu \lambda}\;\mathtt{(20\, equations\, as\, in\, 4D\, GR)}\\ 
R^\nu_{\mu \lambda 5}&=&\alpha F^\nu_{\mu, \lambda}\;\;\;\;\;\;\;\; \;\;\;\;\;\;\;\;\;\;\;\;\;\;\;\;\;\;\;\;\;\;\;\;\;\;\;\;\;\;\;\;\;\;\mathtt{(5\, equations)}\\ 
R^5_{\mu 5 \nu}&=&-\phi_{,\mu\nu}\;\;\;\;\;\;\;\;\;\;\;\;\;\;\;\;\;\;\;\;\;\;\;\;\;\;\;\;\;\;\;\;\;\;\;\;\;\;\;\;\;\; \mathtt{(10\, equations)}\\ 
R^\nu_{\mu 5 5}&=& 0\;\;\;\;\;\;\;\;\;\;\;\;\;\;\;\;\;\;\;\;\;\;\;\;\;\;\;\;\;\;\;\;\;\;\;\;\;\;\;\;\;\;\;\;\;\;\;\;\; \mathtt{(10\, equations)}\\ 
R^5_{\mu 5 5}&=& 0\;\;\;\;\;\;\;\;\;\;\;\;\;\;\;\;\;\;\;\;\;\;\;\;\;\;\;\;\;\;\;\;\;\;\;\;\;\;\;\;\;\;\;\;\;\;\;\;\; \mathtt{(4\, equations)}\\
R^5_{5 5 5}&=&0 \;\;\;\;\;\;\;\;\;\;\;\;\;\;\;\;\;\;\;\;\;\;\;\;\;\;\;\;\;\;\;\;\;\;\;\;\;\;\;\;\;\;\;\;\;\;\;\;\; \mathtt{(1\, equation)}
\end{eqnarray} 
It is clear from the above equations that the curvature of five dimensional metric involves gravitational, electromagnetic and scalar fields.
\subsection{Ricci Tensor}
By successive contraction of the Riemann-Christoffel tensor, we obtain another quantity of great importance, the Ricci tensor,
\begin{eqnarray}
R_{\mu \nu}&=& \Gamma^\lambda_{\mu \nu, \lambda},\;\;\;\;\;\;\;\;\;\;\;\mathtt{(10\, equations\, as\, in\, 4D\, GR)}\\
R_{5 \nu}&=& -\alpha \partial^\mu F_{\mu \nu},\;\;\;\;\mathtt{(4\, equations)}\\
R_{5 5}&=& -\square\phi.\;\;\;\;\;\;\;\;\;\;\;\mathtt{(1\, equation)}
\end{eqnarray} 
All this serves as the left hand side of the field equations. To construct the full field equations we need energy-momentum tensor on the right hand side.
\subsection{Energy-Momentum Tensor and Field Equations}
Now we compute the energy-momentum tensor under the weak field approximation. Then we have, 
\begin{equation}
\hat{T}_{\hat{\mu} \hat{\nu}}=\hat{T}^{\hat{\mu} \hat{\nu}}=\hat{\mu}_0 u^{\hat{\mu}}   u^{\hat{\nu}},
\end{equation} 
 where $\hat{\mu}_0$ is rest mass density and $u^{\hat{\mu}}$ is five velocity. 
\subsubsection{Gravity}
The field equations for gravity are given by, 
\begin{equation}
R_{\mu \nu}= \kappa(T_{\mu \nu}-\frac{1}{2} g_{\mu \nu}T)
\end{equation} 
 from which we can recover the entire general relativity.
\subsubsection{Electromagnetism}
The field equations for electromagnetism are given by, 
\begin{equation}
R_{5 \mu}= -\kappa T_{5 \mu}.
\end{equation} 
From Maxwell's equations, we have 
\begin{equation}
\partial^\mu F_{\mu \nu}= J_\nu=\rho_0 u_\nu.
\end{equation} 
Therefore 
\begin{equation}
\rho_0 v_\nu=({\kappa}/{\alpha})\mu_0 u_5 u_\nu
\end{equation} 
For small velocities i.e.,$u_0=c$ and $u_i \ll c,$ we can write $v_\mu \sim u_\mu$ and setting $\alpha=\sqrt{{\kappa / 2}}$, we get the specific charge as 
\begin{equation}
{\rho_0} / {\mu_0} = 2 \alpha u_5 
\end{equation} 
This equation interprets electric charge as the fifth component of the energy-momentum of matter ``moving across" the $x_5$ space.
\subsubsection{Scalar Field}
Finally, we write the field equations for the scalar field. Because of small velocity approximation, we have $T_{ij}\sim0$, where $i, j$ are only the spatial components. Then, 
\begin{equation}
T=g^{\hat{\mu}\hat{\nu}}T_{\hat{\mu}\hat{\nu}}=-T_{00}=-\mu_0.
\end{equation} 
The field equations for scalar field are give by, 
\begin{eqnarray}
R_{55} &=& \kappa(T_{55}-\frac{1}{2} g_{55}T) \\
R_{55} &=& {{\kappa \mu_0}/2}
\end{eqnarray} 
\subsection{Remarks on Kaluza's Theory}
The Kaluza's approach gave a striking result when working out the Einstein's equations in five-dimensional spacetime, the fifteen higher-dimensional field equations naturally broke into 
\begin{itemize}
\item a set of ten formulae governing a  tensor field representing gravity, 
\item four describing a vector field representing electromagnetism, and 
\item one wave equation for a scalar field.  
\end{itemize} 
Furthermore, if the scalar field was constant,  the vector field equations were just Maxwell's equations in vacuo, and the tensor field equations were the 4-dimensional Einstein field equations sourced by an electromagnetic field. In one fell swoop, Kaluza had written down a single covariant field theory in five dimensions that yielded the four dimensional theories of general relativity and electromagnetism!    

But the following problems plagued Kaluza's theory.
\begin{itemize}
\item{Not the least of which was the nature of the fifth dimension. Before Minkowski, people were aware of time, they just had not thought of it as a dimension. But now, there did not seem to be anything convenient that Kaluza's fifth dimension could be associated with.} 
\item{There was no explanation given for Kaluza's \textit{ad hoc} assumption, that none of the fields in the universe should vary over the extra dimension \textit{(the cylindrical condition)}.} 
\item{This theory could unify gravity with electromagnetism only for small velocity regimes. According Kaluza, a particle's electric charge is related to its velocity component $u^5$. For electron and proton $\rho_o / \mu_0$ is not small and hence $u^5$ should be large. This means, under small velocity approximation, it can explain the macroscopic phenomena but a fundamental problem arises concerning its very applicability to the elementary particles.}
\end{itemize} 
\section{Klein's Modification}
In 1926, Oscar Klein[2] provided an explanation for Kaluza's fifth dimension by proposing it to have a circular topology so that the coordinate $y$ is periodic i.e., $0 \leq y \leq 2\pi R$ where $R$ is the radius of the circle $S^1$. Thus the global space has topology $R^4 \times S^1$.  So Klein suggested that there is a little circle at each point in four-dimensional spacetime. This is the basic idea of Kaluza-Klein compactification. Although there are four space dimensions, one of the space dimensions is compact with a small radius. As a result, in all experiments we could see effects of only four dimensions. Thus Klein made the Kaluza's fifth dimension less artificial by suggesting plausible physical basis for it in compactification of the fifth dimension. The theory of gravity on a compact space-time is called \textit{Kaluza-Klein theory}. 

Klein starts with five dimensional line element (2) and gets the $g_{\mu \nu}, A_\mu$ from $\hat{g}_{\hat{\mu} \hat{\nu}}$ in a similar way as Kaluza did. We set 
\begin{eqnarray}
\hat{g}_{5 5}&=& \phi\\
\hat{g}_{5 \mu}&=& \kappa \phi A_\mu\\
\hat{g}_{\mu \nu}&=&g_{\mu \nu}+  \kappa^2 \phi A_\mu A_\nu
\end{eqnarray} 
Hereby the quantities $\hat{g}_{\hat{\mu} \hat{\nu}}$ are redused to known quantities. Now the new metric can be written as 
\begin{equation}
\hat{g}_{\hat{\mu}\hat{\nu}}=\phi^{-1/3}
\left( \begin{array}{cc}
g_{\mu\nu}+ \kappa^2 \phi A_\mu A_\nu & \kappa \phi A_\mu\\
\kappa \phi A_\nu & \phi\\
\end{array}\right)
\end{equation}
where the field $\phi$ appears as a scaling parameter in the fifth dimension and is called the dilaton field. The fields $g_{\mu\nu}(x,y), A_\mu(x,y)$ and $\phi(x,y)$ transform respectively as a tensor, a vector and a scalar under four-dimensional general coordinate transformations.

The Einstein-Hilbert action for five dimensional gravity can be written as,
\begin{equation}
{\hat S}= \frac{1}{2\hat{k}^2}\int d^5\hat{x} \sqrt{-\hat{g}}\hat{R}
\end{equation} 
where $\hat{k}$ is the five dimensional coupling constant and $\hat{R}$ is the five dimensional curvature invariant. This action is invariant under the five dimensional general coordinate transformations,
\begin{equation}
\delta \hat{g}_{\hat\mu \hat\nu}=\partial_{\hat\mu}\hat\xi^{\hat\rho}\hat{g}_{\hat\rho \hat\nu}+\partial_{\hat\nu}\hat\xi^{\hat\rho}\hat{g}_{\hat\rho \hat\mu}+\hat\xi^{\hat\rho}\partial_{\hat\rho}\hat{g}_{\hat\mu \hat\nu}
\end{equation} 
We can get the field equations of gravity and electromagnetism from the above action by variational principle.

Till now we have re-done the Kaluza's theory. As Klein suggested, the extra dimension has become compact and satisfies the boundary condition
\begin{equation}
y=y+2\pi R,
\end{equation} 
all the fields are periodic in $y$ and may be expanded in a Fourier series,
\begin{eqnarray}
g_{\mu\nu}(x,y)&=& \sum_{n=-\infty}^{+\infty}g_{\mu\nu n}(x) e^{in\cdot y/R}\\
A_{\mu}(x,y)&=& \sum_{n=-\infty}^{+\infty}A_{\mu n}(x) e^{in\cdot y/R}\\
\phi(x,y)&=& \sum_{n=-\infty}^{+\infty}\phi_n(x) e^{in\cdot y/R}
\end{eqnarray} 
with 
\begin{eqnarray}
g^\ast_{\mu\nu n}(x)&=&g_{\mu\nu -n}(x)\\
A^\ast_{\mu n}(x)&=&A_{\mu -n}(x)\\
\phi^\ast_n(x)&=&\phi_{-n}(x)
\end{eqnarray} 

So, the Kaluza-Klein theory describes an infinite number of four-dimensional fields. However, it also describes an infinite number of four-dimensional symmetries since we may also Fourier expand the general coordinate parameter as follows,
\begin{equation}
\hat\xi^{\hat\mu}(x,y)=\sum_{n=-\infty}^{+\infty}\hat\xi^{\hat\mu}(x)e^{inmy}
\end{equation} 
with
\begin{equation}
\hat\xi^{\ast\hat\mu}_n=\hat\xi^{\hat\mu}_{-n}.
\end{equation} 

The equations of motion corresponding to the above action are,
\begin{eqnarray}
(\partial^\mu\partial_\mu - \partial^y\partial_y) g_{\mu\nu}(x,y)&=&(\partial^\mu\partial_\mu + \dfrac{n^2}{R^2}) g_{\mu\nu n}(x)= 0\\
(\partial^\mu\partial_\mu - \partial^y\partial_y) A_{\mu}(x,y)&=&(\partial^\mu\partial_\mu + \dfrac{n^2}{R^2}) A_{\mu n}(x) = 0\\
(\partial^\mu\partial_\mu - \partial^y\partial_y) \phi(x,y)&=&(\partial^\mu\partial_\mu + \dfrac{n^2}{R^2}) \phi_n(x) \;\; = 0
\end{eqnarray} 
Comparing these with the standard Klein-Gordon equation, we get `mass' corresponding to these fields as,
\begin{equation}
m_n\sim \dfrac{n}{R}
\end{equation} 
where $n$ is the mode of excitation. So in four dimensions we shall see all these excited states with mass or momentum $\sim O(n/R)$.  Since we want to unify the electromagnetic interactions with gravity, the natural radius of compactification will be the Planck length,
\begin{equation}
R=\dfrac{1}{M_p}
\end{equation} 
where the Planck mass $M_p \sim 10^{18}GeV.$
\subsection{Graviton Spectrum}
Since the Kaluza-Klein metric is a $5 \times 5$ symmetric tensor, it has 15 independent components. However, because of five dimensional general coordinate invariance, we can impose 5 separate conditions to fix the gauge using harmonic gauge
\begin{equation}
\partial_{\hat{\mu}} \hat{g}^{\hat{\mu}}_{\hat{\nu}}=\dfrac{1}{2}\partial_{\hat{\nu}} {\hat{g}}^{\hat{\mu}}_{\hat{\mu}}.
\end{equation} 
This brings down the number of degrees of freedom to 10. However, this is not yet a complete gauge fixing, the gauge transformations
\begin{equation}
{\hat{g}}^{\hat{\mu}}_{\hat{\mu}}\longrightarrow {\hat{g}}^{\hat{\mu}}_{\hat{\mu}}+ \partial_\mu \epsilon_\nu + \partial_\nu \epsilon_\mu
\end{equation} 
with $\square \epsilon_\nu=0$ are still allowed. This means another 5 conditions can be imposed which results in only 5 independent degrees of freedom. Whereas in four dimensions we have only 2 degrees of freedom for a massless graviton. This implies that from four dimensional point of view a higher dimensional graviton will contain particles other than just ordinary four dimensional graviton.
\subsubsection{Zero Modes}
Now, we study the four dimensional equivalent of the five dimensional action (41), separately for zero and non-zero modes. In what follows we will concentrate on the zero modes of the fields $g_{\mu\nu}(x,y), A_\mu(x,y)$ and $\phi(x,y)$. This is same as dimensional reduction in this particular case. This method of taking the field to be independent of the extra dimension and discarding the non-zero modes is called ``Kaluza-Klein anstaz". 

Substituting (44),(45),(46) and (40) in the five dimensional action (41) and integrating over $y$, we can write the four dimensional action for the $n=0$ modes as, 
\begin{equation}
S=\dfrac{1}{2 k^2}\int dx^4 \sqrt{-g}[R - \dfrac{1}{4}e^{-\sqrt{3}\phi}F_{\mu\nu 0}F^{\mu\nu}_0 -\dfrac{1}{2}\partial_\mu\phi_0\partial^\mu\phi_0]
\end{equation} 
where $2\pi k^2=m\hat{k}^2$ and $F_{\mu\nu 0}=\partial_\mu A_{\nu 0}-\partial_\nu A_{\mu 0}$. This action is invariant under general coordinate transformations with parameter $\xi^\mu_0$, 
\begin{eqnarray}
\delta g_{\mu \nu 0}&=&\partial_{\mu}\xi^{\rho}_0{g}_{\rho \nu 0}+\partial_{\nu}\xi^{\rho}_0{g}_{\rho \mu 0}+\xi^{\rho}_0\partial_{\rho}{g}_{\mu \nu 0}\\
\delta A_{\mu 0}&=&\partial_{\mu}\xi^{\rho}_0A_{\rho 0}+\xi^{\rho}_0\partial_{\rho}A_{\mu 0}\\
\delta \phi_0&=&\xi^{\rho}_0\partial_{\rho}\phi_0,
\end{eqnarray} 
local gauge transformations with parameter $\xi^4_0$
\begin{equation}
\delta A_{\mu 0}=\partial_{\mu}\xi^4_0
\end{equation} 
and global scale transformations with parameter $\lambda$
\begin{eqnarray}
\delta A_{\mu 0}&=&\lambda A_{\mu 0}\\
\delta \phi_0&=&-2 \lambda/\sqrt{3}
\end{eqnarray}
The symmetry of the ground state is  the \textit{four dimensional Poincar\'e group$\times R$,} which is determined by the vacuum expectation values 
\begin{eqnarray}
\langle g_{\mu \nu} \rangle &=& \eta_{\mu \nu}\\
\langle A_{\mu} \rangle &=& 0\\
\langle \phi \rangle &=& \phi_0.
\end{eqnarray}  
Thus, the masslessness of the graviton and photon is due to general covariance and gauge invariance respectively. Whereas the dilation is massless as it is the Goldstone boson associated with the spontaneous breakdown of the global scale invariance. Here the gauge group is $R$ rather $U(1)$, because this truncated $n=0$ theory has lost all the ``memory'' of the periodicity in $y$.
 
Let us count physical degrees of freedom. A four dimensional massless graviton has 2 physical degrees of freedom; a four dimensional massless gauge boson has also 2 physical degrees of freedom and a real scalar has 1 physical degree of freedom
\subsubsection{Non-Zero Modes}
Let us now turn to non-zero modes. The analysis is similar to zero modes of the fields but more combersome. All the above transformations are valid only to the zero modes. Now to determine the properties of non-zero modes we make a restriction
\begin{eqnarray}
\xi^\mu_n(x)&=&a_n^\mu + \omega^\mu_{\nu n} x^\nu\\
\xi^5_n(x)&=&C_n
\end{eqnarray}
where  $a_n^\mu,$ $\omega^\mu_{\nu n}$ and $C_n$ are constants. The corresponding generators are
\begin{eqnarray}
P_n^\mu&=& e^{i n y} \partial^\mu \\
M_n^{\mu \nu}&=& e^{i n y} (x^\mu \partial^\nu-x^\nu \partial^\mu)\\
Q_n &=& i e^{i n y} \partial_y
\end{eqnarray} 
The gauge parameters $\xi^\mu_n$ and $\xi^5_n$  for non-zero modes correspond to spontaneously broken generators which implies that the fields $A_{\mu n}$ and $\phi_n$ are the corresponding Goldstone boson fields. Then the gauge fileds $g_{\mu \nu n}$, with 2 degrees of freedom will acquire mass by absorbing the 2 degrees of freedom from vector Goldstone boson $A_{\mu n}$ and 1 degree of freedom from scalar Goldstone boson $\phi_n$ giving rise to massive spin-2 graviton with 5 degrees of freedom.

From the equation of motion we can say that only the zero modes $(n=0)$ will be massless and observable at our present energy and all the excited states, called as \textit{Kaluza-Klein states,} will have masses 
\begin{equation}
m_n=|n|m
\end{equation} 
as well as charge
\begin{equation}
q_n=n\sqrt{2}\kappa m
\end{equation}  
as shown by Salam and Strathdee[6]. Only when we reach the sufficient energy we can see all these excited states. At the Planck scale we shall also be able to resolve the extra dimension and at even higher energies this fifth dimension will appear to be similar to the other spacial dimensions.                                                 
\section{Conclusion}
Kaluza and Klein's fivedimensional version general relativity although flawed, is an example of such an attempt to unite the forces of nature under one theory. It led to glaring contradictions with experimental data. But some physicists felt that it was on the right track, that it in fact did not incorporate enough extra dimensions! This led to modified versions of Kaluza-Klein theories incorporating numerous and extremely small extra dimensions. The Kaluza-Klein idea of extra spacetime dimensions continues to pervade current attempts to unify the fundamental forces, but in ways some what different from that originally envisaged.
\section*{Acknowledgement}
 We would like to thank Rizwan ul Haq Ansari for useful discussions and many nice dinners!


\begin{thebibliography}{999}
\bibitem{1} Th. Kaluza, Sitz. Preuss. Akad. Wiss. Phys. Math. Kl 996 (1921).
\bibitem{2} O. Klein, Z. F. Physik \textbf{37} 895 (1926); O. Klein, Nature \textbf{118} 516 (1926).
\bibitem{3} G. Nordstr\"om, Phys. Zeitschr \textbf{15} 504 (1914).
\bibitem{4} V.H.Satheesh Kumar and P.K.Suresh, gr-qc/0506125 (2005).
\bibitem{5} M.J. Duff, hep-th/9410046 (1994);\\ M.J. Duff, 280, \textit{An Introduction to Kaluza-Klein Teories,} Ed. by H.C.Lee, World Scientific (1983).
\bibitem{6} A.Salam and J.Strathdee, Ann. Phys. \textbf{141} 316 (1982)
\end{thebibliography}
\end{document}